\documentclass[journal]{IEEEtran}

\usepackage{lineno,hyperref}
\usepackage{tcolorbox}
\usepackage{latexsym}
\usepackage[center]{caption}
\usepackage{multirow}
\usepackage{graphicx}
\usepackage{url}
\usepackage{hyperref}
\usepackage[utf8]{inputenc}
\usepackage{amssymb}
\usepackage{color}
\usepackage{fancybox}
\usepackage{subcaption}
\usepackage{url}
\usepackage{xspace}
\usepackage{amssymb}
\usepackage{multirow}
\usepackage{tabularx}
\graphicspath{{images/}}
\usepackage{url}
\usepackage{tablefootnote}
\usepackage{balance}
\usepackage{graphicx}

\begin{document}

\title{Is Fragmentation a Threat to the Success of the Internet of Things?}

\author{Mohab Aly,~\IEEEmembership{}Foutse Khomh,~\IEEEmembership{}Yann-Ga\"{e}l Gu\'{e}h\'{e}neuc,~\IEEEmembership{}Hironori Washizaki~\IEEEmembership{}and~Soumaya Yacout~\IEEEmembership{}    

\thanks{Manuscript received April 9, 2018; revised July 5, 2018;}
\thanks{Mohab Aly and Foutse Khomh are with the \textit{SWAT Lab.}, Polytechnique Montr\' eal, Montr\' eal,
Quebec, Canada (e-mail: \{mohab.aly, foutse.khomh\}@polymtl.ca).}
\thanks{Yann-Ga\"{e}l Gu\'{e}h\'{e}neuc is with the \textit{Ptidej Team}, Concordia University, Montr\' eal, Quebec, Canada (e-mail: yann-gael.gueheneuc@concordia.ca).}  
\thanks{Hironori Washizaki is with the \textit{Reliable Software Engineering Lab.}, Waseda University, Tokyo, Japan (e-mail: washizaki@waseda.jp).}
\thanks{Soumaya Yacout is with the \textit{MAGI}, Polytechnique Montr\' eal, Montr\' eal, Quebec, Canada (e-mail: soumaya.yacout@polymtl.ca).}%
\thanks{"Copyright (c) 2012 IEEE. Personal use of this material is permitted. However, permission to use this material for any other purposes must be obtained from the IEEE by sending a request to pubs-permissions@ieee.org."}
}


\maketitle

\begin{abstract}

Internet of Things (\textit{IoT}) aims to bring connectivity to almost every objects, i.e., \textit{things}, found in the physical space. It extends connectivity to everyday \textit{things}, however, such increase in the connectivity creates many prominent challenges.

\underline{\textit{Context}}:
Generally, IoT opens the door for new applications for machine-to-machine (\textit{M2M}) and human-to-human communications. The current trend of collaborating, distributed teams through the Internet, mobile communications, and autonomous entities, e.g., robots, is the first phase of the IoT to develop and deliver diverse services and applications. However, such collaborations is threatened by the fragmentation that we witness in the industry nowadays as it brings difficulty to integrate the diverse technologies of the various objects found in IoT systems. Diverse technologies induce interoperability issues while designing and developing various services and applications, hence, limiting the possibility of reusing the data, more specifically, the software (including frameworks, firmware, APIs, user interfaces) as well as of facing issues, like security threats and bugs, when developing new services or applications. Different aspects of handling data collection ranging from discovering smart sensors for data collection, integrating and applying reasoning on them must be available to provide interoperability and flexibility to the diverse objects interacting in the system. However, such approaches are bound to be challenged in future IoT scenarios as they bring substantial performance impairments in settings with the very large number of collaborating devices and technologies. 

\underline{\textit{Objective}}: We raise the awareness of the community about the lack of interoperability among technologies developed for IoT and challenges that their integration poses. We also provide guidelines for researchers and practitioners interested in connecting IoT networks and devices to develop services and applications. 

\underline{\textit{Method}}: We apply the methods advocated by the Evidence-based Software Engineering paradigm (\textit{EBSE}). This paradigm and its core tool, the Systematic Literature Review (\textit{SLR}), were introduced to the sofware-engineering research community early 2004 to help researchers and industry systematically and objectively gather and aggregate evidences about different topics. In this paper, we conduct a SLR of both IoT interoperability issues and the state-of-practice of IoT technologies in the industry, highlighting the integration challenges related to the IoT that have significantly shifted the landscape of Internet-based collaborative services and applications nowadays.  

\underline{\textit{Results}}: Our SLR identifies a number of studies from journals, conferences, and workshops with the highest quality in the field. This SLR reports different trends, including frameworks and technologies, for the IoT for better comprehension of the paradigm and discusses the integration and interoperability challenges across the different layers of this technology while shedding light on the current IoT state-of-practice. It also discusses some future research directions for the community.
\end{abstract}


\IEEEpeerreviewmaketitle

\section{Introduction} 
\label{introduction}

\IEEEPARstart{I}{}nternet of Things (IoT) is a new technology paradigm that is gaining power due to the huge advancements in the electronic and wireless communication technologies fields~\cite{sheng2013survey}. In the coming years, the IoT is expected to bridge diverse Internet collaborative technologies to enable new services and applications by connecting physical objects, \textit{i.e., things}, together in support of intelligent decision making to empower teams across the world. However, this new technology is plagued by interoperability issues that threaten its success. Different categories of interoperability have been emerged; for example, semantic, technical and cross-domain interoperability are from the examples of the interoperability issues that appeared recently in the IoT system. Such kind of interoperability are, hence, needed to advocate seamless, heterogeneous communication in the IoT paradigm. Although it is crucial to provide more technological choices to the end-users, certainties should be maintained; in this context, this means that the IoT requires standards to enable horizontal platforms that are operable, communicable as well as programmable across the participating devices regardless of their manufacturer, model, or industry applications \cite{elkhodr2016internet}.  

The International Telecommunication Union's Telecommunications Standardization sector (ITU-T) has a vision of the IoT as a global infrastructure that enables advanced services by interconnecting physical and virtual \textit{things} based on, existing and evolving, interoperable communication technologies and information \cite{kafle2016internet}, this definition acknowledges M2M as a foundation capacity of the IoT \cite{bassi2008internet, guinard2011internet}. M2M communications are the key enabling technology for IoT where objects communicate with one another for collaborative automation and intelligence optimization; such paradigm features high-quality connectivity to enable ubiquitous messaging as well as interoperable interactions between objects. Recently, a number of studies have highlighted the technologies and frameworks behind IoT, e.g., \cite{gazis2017survey, li2015internet, kohler2014platforms}, however, they discussed these technologies in isolation and did not examine issues related to their integration. 

The vision of IoT is couched on wide scale interoperability among multiple domains; this brings on technical challenges in the area of data dissemination mechanisms, data representation formats and data management platforms. Those must collectively support the integration of various types of data generated from disparate sources that are possibly operating under different administration strategies to enable flexible data mashups, hence, foster rapid innovation in the application and services ecosystems \cite{ETSI1, ETSI2}. It becomes apparent that most of the challenges brought by IoT are that ones related to interoperability concerning multiple layers of the end-to-end protocol stack. Then, standardization of key interoperability areas has been recognized consistently as the most significant factors for the success of the entire IoT system and its economy \cite{ETSI3}

This paper aims to raise the awareness of the community about the lack of interoperability among technologies developed for IoT and challenges that their integration poses. The remainder of this paper is organized as follows, Section~\ref{background} depicts the background of the IoT. Section~\ref{methodology} describes the methodology followed in this review. Section~\ref{interoperabilty challenges} describes the possible integration and interoperability challenges against the IoT paradigm. Sections~\ref{solutions} and \ref{state of practice} summarize the solutions against those interoperability issues presented in Section~\ref{interoperabilty challenges}. In Section~\ref{future} we talk about the future visions of IoT deployments that need to be considered accordingly, and finally, Section~\ref{conclusion} presents our conclusion and outlines the main findings of this review.

\section{Background on the Internet of Things} 
\label{background} 

There is no exact definition of IoT yet since it is still in the forming process and is subject to the perspectives taken \cite{hepp2007harvesting,joshi2008survey,pretz2013next}. It was first introduced to the community as a ``dynamic global network infrastructure with self-configuring capabilities based on standards and interoperable communication protocols''. In IoT, physical and virtual \textit{things} have identities and attributes that are capable of using intelligent interfaces and being integrated as an information network \cite{kiritsis2011closed,li2012effects,li2012efficient}. This concept can be seen as a superset of connecting devices that are identifiable in a unique way by the help of existing near field communications (NFC) techniques \cite{kohler2014platforms,key2013}. When the two words ``Internet'' and ``\textit{Things}'' combine together, this implies an interconnected world-wide network based on infrastructure, communication, networking and information processing technologies \cite{nic2008disruptive}.

The sensory capabilities of devices have been extended significantly via the emerging wirelessly sensory technology, thus the original concept of IoT is therefore extending to the ambient intelligent and autonomous control. Multiple technologies are involved in the IoT paradigm, such as wireless sensor networks (WSNs), Cloud Computing, RFID, NFC, low energy wireless communications, etc. \cite{jiang2014iot,kataev7enterprise,li2013applications,ren2012methodology,tao2014cciot,wang2014iot}. With the evolution of these technologies, new technologies are brought to the IoT framework \cite{deng2010new,van2011internet,li2012integration,malatras2008web,miorandi2012internet,pautasso2009web,peris2006m,wang2012advances}. Physical \textit{things} could be identified, accessed and operated via the Internet technology. Depending on various mechanisms for the implementation, the definition of IoT and its context could vary. 

With the advancement of cellular networks and mobile phones, telemetry evolved to the newly concept of M2M communications. Applications are ranging from fleet management, asset tracking, etc; everyday objects (e.g. cars) and the environment (e.g. parking lots) get network connectivity to enable interactions. The term IoT and M2M are interchangeable in the industry, especially in the network provider sector. In contrast, today's M2M applications are heavyweight custom solutions that are in no way based on interoperability or standards; notably, M2M is a business-centered term whereas the vision of ubiquitous computing is a human-centered one. In conventional M2M scenarios, computations and intelligence are centered at a specific location (i.e. central enterprise applications in an RFID scenario) whereas the ubiquitous computing cares about distributive computing where the computational tasks can be divided and distributed among the pre-configured network \cite{kohler2014platforms}. Related to the back-then formed ubiquitous computing technology, in the \textit{mid-1990's}, the wireless sensor networks (WSNs) and wireless personal area networks (WPANs) fields emerged; both build self-organizing networks of smart objects that can interact very well with the physical surroundings \cite{dargie2010fundamentals}. Typical WSN/WPAN scenarios involve proprietary protocols, such as Z-Wave and ZigBee, where gateways are put in place to translate the formats between proprietary and IP networks. Furthermore, there are small WPAN solutions that use Bluetooth where smart phones conventionally act as a gateway or mediator to facilitate objects' interaction (e.g. smart watches). The potentiality of IoT services and applications in various fields boosts the study of a wide spectrum of relevant techniques. Flexible connectivity, ubiquitous messaging and interactive collaboration technologies are the promising trends of the IoT computing paradigm; they allow greater efficiency and intelligent optimizations among the participating \textit{things} (i.e., devices) based on the insights of the sensory data collected. The emergence of IoT is transforming the way we communicate. Connected entities are now integrated to the Internet, allowing us to process and consume feeds of data collected from the field in real time, and interact with each other in virtual environments. 

In the industrial sectors nowadays, machines are equipped with sensors to help monitor their health and communicate important information about their status to specialized teams distributed around the world, allowing the integration of such information in real time communication to make decisions about the working systems. The next step for those industrial machines is to be integrated in a virtual environment; this enables not only realistic representation of the past and present state of the participating machines, but also forecast of possible future scenarios. End-to-end interoperability should be considered to ensure the proper delivery of services regardless of the specifications of the machines used.

\begin{figure*}[ht] 
\centering
\includegraphics[width=0.7\linewidth]{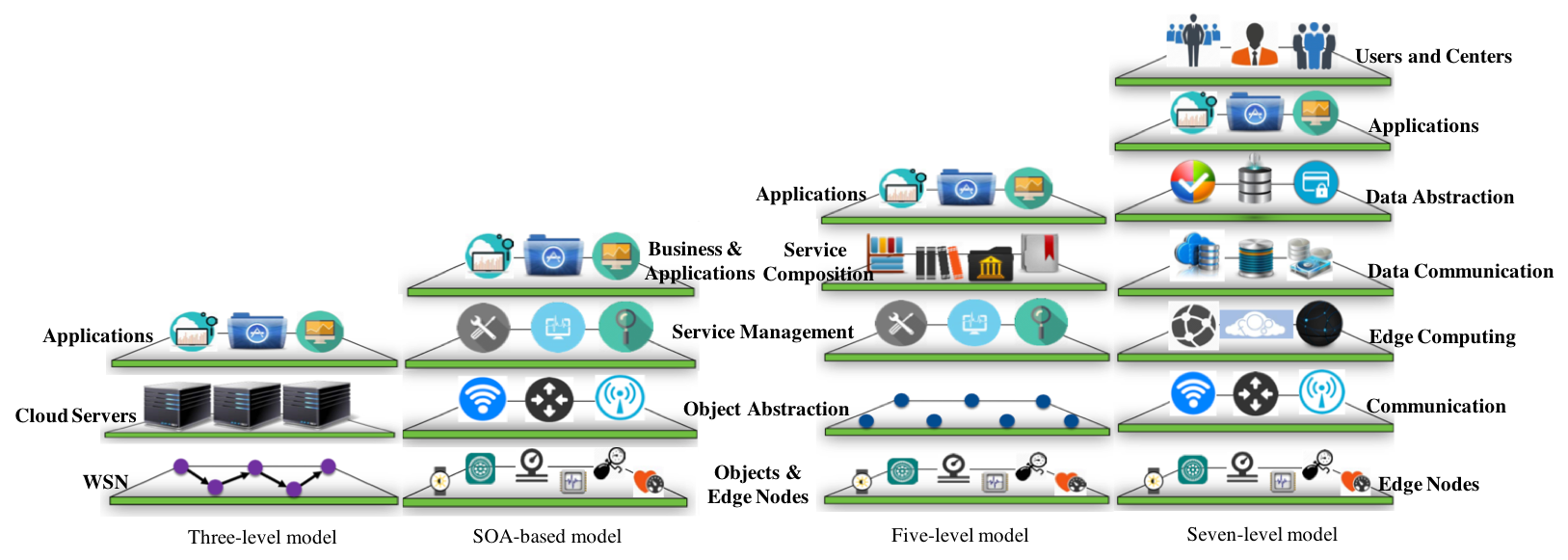}
\caption{Three, SOA-based, Five and Seven-level Internet of Things reference models}
\vspace{-0.3cm}
\label{figure1}
\end{figure*}

\subsection{Architectures for the Internet of Things}

The highly competitive nature of the IoT makes interoperability between different \textit{things} a difficult task to achieve. A crucial need of an IoT ecosystem is that \textit{things} found within the network must be interconnected to exchange information. System architecture should be able to guarantee the practical operations of the IoT to bridge the gap between the physical and logical (i.e., virtual) worlds. IoT is characterized by network architectures that are not necessarily fixed, but highly dynamic; thus, having consequences on applications that need services that could once have been available and no longer are and on services that might depend on devices that had once joined the network and left suddenly (temporarily or permanently) or may need devices that never existed in the targeted location. 

Architectures should be flexible and adaptive to let smart \textit{things} interact with one another in a dynamic way, they also need to support the unambiguous communication of the triggered events; this is due to the fact that smart \textit{things} have the ability to move geographically to get involved with others in real-time mode. The ever elevating number of suggested architectures has not converged to a reference model yet \cite{krco2014designing}. There is no universally-accepted overall IoT architectural framework although there are several efforts underway to achieve some convergence. 

Several IoT architectural models have been advanced in recent years, \textit{figure \ref{figure1}} \footnote{adopted from:~\cite{nia2017comprehensive}}, each of them is focusing on some specific formulations or abstractions of the IoT ecosystem \cite{weyrich2016reference}. The \textit{three-level model} \cite{gubbi2013internet} is among the proposed architectures; it depicts IoT as an extended version of wireless sensor networks (WSNs) and models it as a combination of Cloud servers and WSNs which offers different services to users. The service-oriented architecture (SoA)-based architecture ensures the interoperability among heterogeneous smart devices in multiple ways \cite{panetto2013information,jardim2013systematisation,wang2012dimp}; it has been suggested to add more abstraction to the IoT architecture \cite{khan2012future,yang2011study,chaqfeh2012challenges}. The \textit{five-level model} \cite{atzori2010internet, al2015internet}, an alternative to the \textit{three and SOA-based} ones, has been proposed to facilitate the interactions between different sections of an enterprise by decomposing complex systems/services into simplified applications consisting of an ecosystem of simpler and well-defined components. 

\begin{table*}[ht]
\centering
\caption{Seven-level Internet of Things model} 
\begin{tabular}{ p{2.5cm}|p{15cm} }
\hline
\textbf{} & \textbf{\footnotesize Description} \\
\hline
\footnotesize Collaboration and processes & \footnotesize - highest level of IoT model where users reside; users make use of the applications and the analytical data therein. \\ \cline{2-2}
\footnotesize Applications & \footnotesize - this level provides ``information interpretation'' where software cooperates with data accumulation and data abstraction levels. IoT applications are numerous and might vary across markets and industrial needs. \\ \cline{2-2}
\footnotesize Data abstraction & \footnotesize - the opportunity to render and store data is provisioned in this layer; hence further processing becomes easier and more efficient. Common tasks of entities at this level include normalization, denormalization, indexing, consolidating data into one place and providing access to multiple datastores. \\ \cline{2-2}
\footnotesize Data accumulation & \footnotesize - not all the applications need instant processing; this level enables conversion of data in motion to data at rest, i.e., it allows the storage of data for future analysis or to share with high-level computing servers. Converting the format from network packets to database tables, reducing data via filtering and selective storing and determining whether the data are of interest to higher levels are of the main tasks of this level. \\ \cline{2-2}
\footnotesize Edge Computing & \footnotesize - this is a level in the model where simple data processing is initiated; this is essential for reducing the computational load in higher levels and providing a fast response as well. Nowadays, most real-time applications need to perform computations as close as possible to the edge of the network; the amount of processing in this level depends on the computing power of each of the servers, computing nodes as well as service providers. Typically, simple signal processing and learning algorithms are utilized in-here. (AKA Fog Computing) \\ \cline{2-2}
\footnotesize Connectivity & \footnotesize - consists of all the parts and components that enable the transmission of information and commands: (a) communication between devices; (b) communication between the components, and; (c) transmission of the information between the edge devices and computing (edge computing levels). \\ \cline{2-2}
\footnotesize Edge devices and controllers & \footnotesize - \textit{consists of computing nodes, such as RFID readers, sensors, smart controllers and different versions of RFID tags. Both data confidentiality and integrity must be taken into account from this level upwards.} \\ 
\hline
\end{tabular}
\label{table:SevenLevelModel}
\end{table*}

Recently, in \textit{2014} and \textit{2017} respectively, CISCO and the Open System IoT Reference Model (OSiRM) suggested a comprehensive extension to the traditional \textit{three}, \textit{four} and \textit{five}-levels models by introducing their \textit{seven-level models} that have the potential to be standardized, hence, creating a widely accepted reference model for the IoT paradigm \cite{minoli2017iot}. In such models, data flow is bidirectional in nature, however, the dominant direction of the flow of data depends on the application being worked on. For example, in a control system, data and commands travel from the top of the model (the applications level) to the bottom (edge-node level), whereas in a monitoring scenario, the flow is vice versa (from bottom to top). Providing detailed descriptions of all aforementioned four IoT reference models is beyond the scope of this paper. To summarize IoT recent technologies and frameworks in a level-by-level fashion, we consider the \textit{\textbf{CISCO}} reference model in this paper since it particularly summarizes the up-to-date modeling approach for IoT. We briefly describe each level of this model in table (i.e., \textit{Table \ref{table:SevenLevelModel}}).

Allowing devices for data interaction and cross-platform interoperable communications is an important step towards device collaboration in the digital world. However, there are multi-fold meanings for \textit{things} communications about hierarchical architecture of IoT-ized systems, including discovery and connectivity, messaging systems and mechanisms, and semantic interoperability. The later factor of \textit{things} communications is achieved via seamless integration of the underlaying protocols and standards. 

\section{Prior and Related Works}

Most popular technologies in IoT, e.g., sensor network technologies, have been the subject of other surveys. Contributions in regards to IoT protocol standardization has been summarized by Sheng et al.\ \cite{sheng2013survey}. However, this previous work focused only on some specific technologies (e.g., the IEEE 802.15.4 standard as well as the details of 6LowPAN work) and did not investigate interoperability issues.

The essential features and the key capabilities of wireless protocol stack for IoT is, comprehensively, addressing vital requirements in reliability, efficiency, and connectivity to the Internet, is surveyed by Palattella et al.\ \cite{palattella2013standardized}. Although that work presents some key technologies for the IoT in the wireless domain, focusing on the IEEE 802.15.4 protocol stack and its pertaining features in support of reliability, efficiency and Internet connectivity. However, other key technologies for the IoT (e.g., capabilities of each layer, etc.) are not addressed, hence, providing only a narrow perspective into IoT standardizations (as opposed to the scope of this paper). Another work by Gubbi et al.\ \cite{gubbi2013internet} surveys the potential applications, technological drivers, challenges, and future research areas for the IoT, but relevant activities in each of standardization and--or interoperability are neither identified nor discussed.

Al et al.\ \cite{al2015internet} addresses the technological pillars of the IoT, primarily from the research perspective of: (a) the technologies employed in wireless communications among a group of IoT nodes, and (b) the interoperable data exchange between the IoT and Internet nodes. Yet, the need for stronger horizontal integration at the IoT above layers is identified but the contributions of relevant standards addressing the interoperability between nodes are not discussed.

\section{Study Methodology} 
\label{methodology}

The following subsections present our proposed methodology to perform the Systematic Literature Review (SLR), and the outcomes of the SLR:

\subsection{Conducting the Study}
\subsubsection{Data Sources}

\textbf{Literature collection} was done by making a comprehensive systematic search on the major indexing databases following the guidelines given by \cite{kitchenham2009systematic}. We used \textit{ACM digital library}, \textit{ScienceDirect}, \textit{Springer}, \textit{IEEE Xplore}, \textit{Engineering Village}, \textit{Web of Science} and \textit{Google Scholar} and did an electronically-based search considering the following terms: 'Internet of Things' AND 'interoperability' AND 'integration' AND 'architecture' AND ('platform' OR 'models' OR 'technology' OR 'framework' OR 'trend' OR 'protocols' OR 'standards') AND 'future directions'. We researched for the published scientific papers related to the IoT technologies between the years of \textit{2000} and \textit{2017}, then we did constraint our study to a number of journals, conferences and white papers which having the highest quality in their fields. We carried out this step by choosing the published studies in journals, conferences with high impact factor and competitive acceptance rate. We also check the citation count of the studies being chosen on \textit{Google Scholar} to evaluate their impact on the evolution of this emerging paradigm. Other studies are excluded for quality reasons (e.g., the study is only a small increment over a previous study, a technical report that is extended into a journal or a conference/workshop paper, etc.), however, if a conference paper is extended into a journal version, we only consider the journal version out of it. We excluded the studies that are not published by well-known publishers or did not pass through the well-defined referring processes as explained by \cite{kitchenham2009systematic,kuhrmann2017pragmatic}.

\textbf{To gather information about the state-of-practice of IoT}, we searched the websites of the major hands-on technology providers and downloaded the white papers published to get a grasp of them, also we searched for the tech blogs published by those technology leaders to identify the challenges in the implementation of IoT technologies. Such blogs provide up-to-date information news and all the technical aspects needed to dive deep into the fundamentals of IoT; they cover various aspects ranging from technical point of views to use cases and white papers. 

\subsection{Search and Selection Processes}

Relevant studies from the aforementioned data sources are organized in three rounds as described in Figure 2.

\begin{itemize}
\item \underline{\textit{Round 1}}: We perform electronic search and we narrow our scope review to identify and categorize the preliminary studies related to our subject, i.e., integration and interoperability. Then, we read and select the most relevant studies based on their titles and abstracts; any irrelevant studies are removed. 
\item \underline{\textit{Round 2}}: We read the remaining studies very carefully then any irrelevant studies are eliminated based on the selection criteria identified in the work of \cite{dybaa2008empirical}, we apply different inclusion and exclusion criteria on the remaining studies. These selection criteria can help decide whether to include a paper for further search. Only relevant studies that are retained will be used in this paper analysis. (1) Only papers describing issues related to IoT interoperability and integration are included. (2) Documents presented in the form of powerpoint presentations, abstracts and submitted papers are not included within this study. 
\item \underline{\textit{Round 3}}: Following existing guidelines \cite{wohlin2014guidelinesMac,kuhrmann2017pragmatic}, we perform a snowball search using the reference list of our studies and citations obtained from the previous Round 2 to identify new studies and decide whether to include additional paper(s) or article(s); such technique helps us not to miss important and relevant papers related to the field. Those remaining papers are read carefully afterwards. 
\end{itemize}

One way to narrow down the search space is to conduct a preliminary investigation of the field, i.e., integration and interoperability issues in IoT, by relying on snowballing. The investigation starts by studying publications known in advance and by iteratively extending the known literature set by following the references provided therein \cite{kitchenham2015evidence}. This procedure helps to provide an overview of the publication space and key contributors to conduct the review.

\subsection{Quality of the Selected Papers}

We apply various inclusion and exclusion criteria on the remaining set of studies that resulted from the second and third rounds. These selection criteria help to decide whether to include a paper for further investigation. Below are the criteria used in our SLR. 
\begin{itemize}
\item [-] Documents in form of abstracts, powerpoint presentations or abstracts are excluded. 
\item [-] Papers touching issues related to IoT interoperability, fragmentation as well as proposing solutions to address these issues are included. 
\end{itemize}

\subsection{Organization of the SLR}

The following subparagraphs describe the motivation behind the following tackled parts in this SLR:

\textbf{\textit{Part 1}}: Integration and Interoperability challenges in IoT: this part describes the potential interoperability issues that can affect the IoT paradigm. Such study aims to provide a comprehensive overview of common integration and interoperability issues and challenges that hinder the IoT systems. Furthermore, it can help researchers aiming to improve interoperability in IoT identify future research directions. 

\textbf{\textit{Part 2}}: Interoperability Solutions in IoT: this part reviews the solutions and countermeasures proposed to improve IoT's integration and interoperability.

\textbf{\textit{Part 3}}: Software development issues in IoT: this part sheds the light on software architecture and solutions to be followed to mitigate the interoperability issues identified in this SLR. It draws a roadmap for further studies in IoT software development.

\textbf{\textit{Part 4}}: Research future directions: this part introduces some future research directions that can be considered to cover additional integration and interoperability issues in the IoT ecosystem.

\begin{figure}[t]
\centering\includegraphics[width=0.5\textwidth]{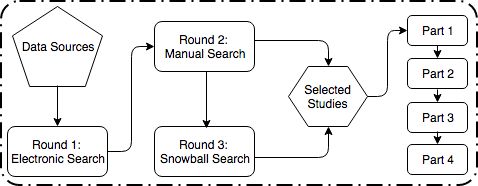}
\centering\caption{Overview of the SLR Methodology}
\end{figure} \label{fig:2}

\section{Integration and Interoperability Challenges} 
\label{interoperabilty challenges}

IoT integration has become a problem due to of its expensiveness \cite{key2017_1}. Not only that, but also it is hard to keep IoT physical parts up-to-date as all devices depend on an integration to provide access or information, hence, can span a wide diversity of technologies, locations, operations and sensitivity levels. The data that the devices provide is usually vast in nature, but might be hard to be transmitted because of the physical limitations of the contributed devices and their environments. 

As the IoT evolves, future networks will continue to be heterogeneous, manufactured by multi-vendors, providing multi-services and will be largely distributed. As a consequence, the risk of non-interoperability will increase; this might lead to the unavailability of some of the provisioned services for end-users who can have harmful consequences with regards to the applications related, for example, to emergency health, etc. Or it could also mean that users/applications are likely to loose key information resulted out of IoT due to this lack of interoperability. Hence, it is important to ensure that network components will interoperate to unleash the full value of the IoT paradigm.

Interoperability is considered a key challenge in the realms of the IoT. This fact resides true due to the intrinsic fabric of the IoT as: (1) highly-heterogeneous, where vast systems are conceived by lots of manufactures and are designed for various purposes targeting variety of application domains, making it difficult (if not impossible) to reach out for global service agreements and widely accepted specifications; (2) high-dimensional, with the co-existence/collaboration of different systems (i.e., sensors, devices, machines, etc.) in an environment that relies on communication and exchanging of information; (3) dynamic and non-linear, where new \textit{things} (that were not even considered at start) are able to join (and leave) the environment at any time and that support new unforeseen formats and protocols, but they need to be able to communicate and share data in the IoT paradigm; and (4) the hardness to describe/model due to the presence of different formats, described in various languages, that can or not share the same modeling principles. This qualifies interoperability in the IoT as a problem of complex nature. We therefore need approaches and comprehension of Interoperability for the IoT also making sure it endures, that is sustainable by discussing the protocols and standards that help achieving such task seamlessly.  

\subsection{Network-layer Interoperability}

Power constrained devices require efficient networking standards and protocols. Conventionally, the paradigm is scattered between a number of different power networking protocols (e.g., ZigBee, Bluetooth), traditional networking protocols, like Ethernet, WiFi, as well as hardwired connections. Such protocols are suggested for domain specific applications that have the ability of provisioning of distinctive features. Solving interoperability issues at this stage requires standardizations at the software and hardware levels. Different products have been developed to support a number of networking protocols by grouping the required software and hardware components together.   

Services in IoT require wireless communications to deploy IoT smart devices easily. As a consequence, routing and communication protocols are considered to be important function to realize practical wireless networks; Dynamic Source Routing (DSR) \cite{johnson2007dynamic} and Adhoc On-Demand Distance Vector (AODV) \cite{perkins2013dynamic} are the renowned routing protocols to be used within the adhoc networks, also a geographic routing is suitable for unstable networks as Vehicular Adhoc Networks VANETs \cite{jerbi2009towards}. 6LowPAN makes an assumption that the IPv6 Routing Protocol for Low power and Lossy Networks (RPL) is a routing protocol in the IoT paradigm \cite{clausen2011critical, pavkovic2011multipath,ko2011evaluating}, thus, an implementation and evaluation software have been developed on real devices \cite{cody2009performance,dawans2012link,baccelli2011p2p,oikonomou2012stateless,hammerseth2011implementing,saad2011simulation,herberg2011comparative}. Services in IoT rely on upper layer protocols, such as an application layer \cite{bormann2012coap,chen2012rpl}, some research efforts have developed CoAP on real IoT devices to see its impact on them \cite{kuladinithi2011implementation,becker2011deployment}. With respect to communications level, various optional protocols, middlware and applications programming interfaces (APIs) libraries are promising for M2M messaging. Despite the fact that they are based on various different techniques for different application scenarios, the goal is to achieve flexibility and interoperability in interactions among participating devices. 

\subsection{Messaging-protocol Interoperability}

In newly developed IoT applications, a number of application level protocols, see Section~\ref{solutions}, are proposed by different enterprises to become the de-facto standards to help the provisioning of communication interoperability \cite{desai2015semantic}. Each protocol possesses specific messaging architecture and unique characteristics that are helpful for various types of IoT applications, which require efficient utilization of the limited energy and processing power capabilities. However, the scalability nature of IoT architecture needs to be independent of the messaging protocol standards, besides providing translation and integration between different popular messaging protocols. 

\subsection{Data Annotation-level Interoperability}

Conventional IoT service model provides raw data captured from the heterogeneous collaborated \textit{things} found within the system. Such data do not contain intellectual annotation that needs extensive manual efforts to build practical usable applications. Because of the proprietary approaches employed by the IoT providers, the IoT system has switched to a domain of vertical compartments of different applications with no proper horizontal connectivity among them. This lacks of interoperability with self-dependent services endangers the acceptability and adoption of the IoT domains, specially for the applications that gain benefits from the number of different participating devices.    

\begin{table*}[ht]
\centering
\caption{Relationship between M2M and IoT Standards} 
\begin{tabular}{ p{1.3cm}|p{2.0cm}|p{13.8cm} }
\hline
\textbf{\footnotesize Standard} & \textbf{\footnotesize IoT Layer} & \textbf{\footnotesize Description} \\
\hline
\footnotesize OneM2M & \footnotesize \footnotesize Service & \footnotesize - Defines a common set of capabilities to support M2M applications, respective access interfaces and the protocols employed over such interfaces, without restricting the technological solutions that could be employed to achieve such capabilities. \\ \cline{3-3}
\footnotesize ETSI & \footnotesize \footnotesize Service & \footnotesize - Defines a common set of capabilities to support M2M applications and the reference points at which such capabilities are accessed independently of the instrumenting technological solutions. \\ \cline{3-3}
\footnotesize ITU-T & \footnotesize \footnotesize Service & \footnotesize - Defines a common set of capabilities to support M2M applications and the reference points at which such capabilities are accessed independently of the instrumenting technological solutions.  \\ \cline{3-3}
\footnotesize OASIS & \footnotesize \footnotesize Data & \footnotesize - Defines generic and flexible mechanisms for defining identity information for \textit{things} and exchanging identity information between different administrative domains. \\ \cline{3-3}
\footnotesize IEEE & \footnotesize \footnotesize Communication & \footnotesize - Architecture harmonization and multiple application domains support (i.e., verticals). \\ \cline{3-3}
\footnotesize IETF & \footnotesize \footnotesize Communication & \footnotesize - Application guidelines provisioning to fit the operation of specific protocols in an IoT setting. Also, defines additional protocols to fill gaps in the protocol solution sets for IoT. \\ 
\hline
\end{tabular}
\label{table:IoTLayeredStandards}
\end{table*}

\section{Proposed Solutions to Integration and Interoperability Issues} 
\label{solutions}

To address the above challenges, the research community and industry have been working on the development of standard and implementation practices that would allow a better communication between the services provided by different providers and help ease the pain of their integration. 

In the following, we discuss some key standards IoT technologies along the different layers of the IoT architecture and outline proposed integration guidelines.

\subsection{Standards and Technologies}

IoT requires a number of different technologies. Specially, communication technologies are considered to be a fundamental framework to realize various IoT services. Standards, on one hand, help both developers and users to determine the best technical protocol for dynamic services and applications in IoT. On the other hand, standardization of technologies is crucial that can and will accelerate the speed of the IoT technology. 

Recently, some efforts have been put in place to make the incorporation of IP protocol stack into smart object possible. The IP stack has to be adapted since the requirements of smart objects differ than that of the usual participant of Internet nowadays \cite{rodrigues2010survey}. Such incorporation has to be done in such a way that a transparent end-to-end connection between devices over the Internet is achieved. To achieve such purpose, a number of protocols have been standardized by the Internet Engineering Task Force (IETF), such as CoAP (constrained application protocol), 6loWPAN (IPv6 for low power wireless personal networks) and ROLL (routing over lossy links). On the other side, i.e. M2M, standardization processes that are driven by the European Telecommunications Standards Institute (ETSI) prepared the Long Term Evolution (LTE) networks for massive low-throughput non-human communications, therefore, full IP connectivity in individual things. Standards play a vital role for further developing and spreading of IoT services; they aims at lowering the entry barriers for both new service providers and users. This will improve the interoperability of different systems/applications and allow products/services to perform better at higher levels \cite{miorandi2012internet}.

IoT standards have attracted research communities attention to its development \cite{li2015internet}; internationally, Electronic Product Code global (EPCglobal), the ITU, International Elector-technical Commission (IEC), International Organizations for Standardization (ISO) and IEEE provided a number of standards to make the identification, capturing and sharing data using RFID technologies easy. On the other hand, the European Telecommunications Standards Institute (ETSI) and the European Committee for Electro-technical Standardization (CEN/CENELEC) released a set standards on the IoT fundamental technologies, such as WSNs, RFID, etc. Moreover, the American National Standards Institute (ANSI) in the United States is proactively working on the management standards of IoT. It is worth to stress on the importance of standards for the IoT technological development. Not only standards help to determine the best technical protocols to be used for dynamic applications and service in IoT, but also it is important where they can help in accelerating the spread of the IoT technology.

M2M has a similar meaning of IoT in its context \cite{kim2014m2m}; many authors consider that M2M is focusing on the automatic cooperation between participating entities comparing to IoT to achieve desired services. Although typical M2M devices are not equipped with enough computation power due to hardware specifications or limitations, simplified protocols for resource constrained devices have been suggested. Industry 4.0 \cite{kagermann2011industrie, kagermann2014industrie, drath2014industrie} and Industrial Internet Consortium (IIC) \cite{industrial2014introductory, industrial2015industrial} have considered standards for practical applications on M2M in an industrial domain; they are proactively trying to develop new application platforms for manufacturing floors, such as factories, manufacturing facilities, etc. In the future, direct communication mechanisms among different M2M devices are required to realize flexible and scalable service systems among a number of service domains \cite{sabella2016mobile}. Table~\ref{table:IoTLayeredStandards} summarizes the relationship between M2M and IoT standards in a layered fashion.  

The IoT puts into consideration both constrained nodes and networks, therefore, historical full stack protocols are not adequate to be deployed on constrained nodes. As a consequence, constrained protocols have been suggested for an IP network as well as the application layers.  

\begin{itemize}
\item \textbf{Message Queue Telemetry Transport} (\textit{MQTT}) \cite{locke2010mqtt} is a M2M/IoT connectivity protocol that is extremely lightweight publish/subscribe messaging transport; it is useful for connections with remote locations where a small code footprint is required and/or network bandwidth is at a premium. Multiple clients connect to a broker then subscribe to one of the present topics. When clients are connected to that specific broker, they are able to publish their messages to the topic(s) they are subscribed to. Since topics are seen as hierarchy in their nature, clients are able to handle all topics in the same way as a file system. The protocol defines three Quality of Service (QoS) levels: (a) \textit{QoS 0} that is related to delivering a message once with no confirmation, (b) \textit{QoS 1} for delivering a message at least once with confirmation required, and, (c) \textit{QoS 2} related to delivering a message exactly once by using a four-step-handshake. With those levels, clients and publishers can control the QoS delivery levels according to the service model being considered. Since MQTT protocol requires an underlying transport that provides an order and reliable communications, TCP is exclusively used for MQTT to fulfill such necessity. Additionally, TLS is being used to realize a secure function on top of the MQTT protocol. 

\item \textbf{Constrained Application Protocol} (\textit{CoAP}) \cite{shelby2014constrained} is a simple application layer protocol that is used by simple electronic devices. It enables such nodes to communicate with wider Internet using similar protocols. Traditionally, CoAP is designed to easily translate different format to Hypertext Transfer Protocol (HTTP) for simplified integration with web systems while also meeting some specialized requirements, such as simplicity, very low overhead and multicast support \cite{colitti2011evaluation, colitti2011integrating}. It provides a request/response interaction model between application endpoints, hence, proxying between CoAP and HTTP can easily have messages translated through an intermediary. Simplicity, very low overhead and multicast are crucial for the IoT and M2M devices that tend to be deeply embedded and have very low memory and power supply; in that essence, efficiency is very important factor. CoAP has the ability to run on most devices that support User Datagram Protocol (UDP) or UDP analogue, optionally to Datagram Transport Layer Security (DTLS), to provide a high level of communication security. Since it defines two types of messages (i.e., request and response), it uses two simple types of messages, named requests and responses. The format of the header (short fixed-length binary 4 bytes) is shared by these types of messages where each message contains message ID which is used to detect duplicates. In CoAP, the message procedures are carried with either a method or a response codes, respectively.

\item \textbf{IPv6 over Low power Wireless Personal Area Networks} (\textit{6LowPAN}) \cite{mulligan20076lowpan, culler20096lowpan} is designed in favor of low-power devices with limited processing power capabilities. The 6LowPAN group defined encapsulation and header compression mechanisms so that the IPv6 packets can be sent and received over the IEEE 802.15.4 base networks. It provides some functionalities, such as adaption layer for interoperability and packet formats between IPv6 and IEEE 802.15.4 domains, address resolution between IPv6 addresses and IEEE 64 bit extended addresses on IEEE 802.15.4 and adapting packet sizes between a traditional IP network and an IEEE 802.15.4 network. Maximum message size is limited up to 128 bytes in the IEEE 802.15.4, hence, the fragmentation process associated with this protocol is optimized to covey an IPv6 payload effectively. The protocol uses a dispatch field that is found in the first part of the packet to recognize a type of the packet and defines two types of dispatches as well, first and subsequent fragments, for carrying an IP datagram. Conventionally, the first fragment is used to carry a compressed IPv6 header information, a transport layer header and a first part of a payload; on the other hand, the subsequent fragments are used to carry only a part of the IPv6 datagram payload since the compressed IPv6 header should be an overhead in the limited payload size of IEEE 802.15.4.
\end{itemize}

\subsubsection{\textbf{Wireless Local Area Network}}

IoT embedded devices use a communication facility to connect to the Internet. A well-known network standard, IEEE 802.3 (the Ethernet) \cite{law2013evolution, keyieeewlan}, is used to provision such functionality. Some of the IoT devices do employ such standard to connect to a network when the participating devices are fixed in a facility because power over Ethernet (PoE) can provide electric power to devices as well. Lately, the power usage of IoT devices has been rapidly reduced according to the advancement of the semiconductor technology. As a consequence, modern IoT devices use wireless communication devices to get access to the Internet. Below is the main stream standards to achieve a local network. 

\begin{itemize}
\item \textbf{IEEE 802.11} \cite{mangoldieee, keyieee802.11} is a known standard for Wireless LAN; recently, WLAN devices support IEEE 802.11n/ac that support Multi-Input Multi-Output (MIMO) technologies on both 2.4 and 5 GHz bands. New standards for different bands have been suggested because existing wireless bands for WLAN are limited and crowded.

\item \textbf{IEEE 802.11ad} is a Wireless Gigabit (WiGig) standard which supports 7 Gbps on 60 GHz band. The purpose out of it is to enable the provisioning of a high throughput performance in a limited space; this is because a 60 GHz signal attenuates proportionally with the distance.

\item \textbf{IEEE 802.11af} is a similar Orthogonal Frequency Division Multiplexing (OFDM) technology on TeleVision White Space frequency spectrum (TVWS) which is sub-GHz bands. The OFDM signal's bandwidth in such standard is 6 or 7 MHzs' that is equivalent to the TV broadcasting signal. IETF provides a standard for channel sensing method since the TV towers is using TVWS bands; conventionally, IoT devices should have the ability to sense signals from TV towers or check TVWS databases for adoption. IEEE 802.11af is becoming the main stream standard that provides high throughput performance on TVWS sub-GHz bands.

\item \textbf{IEEE 802.11ah} resembles IEEE 802.11af in its functionality as both standards make use of the TVWS bands available. Despite of their similarity, the target of each one differs than from the other; the IEEE 802.11ah focuses on long-distance communication with low power consumption and supports 1 and 2 MHzs' bands to increase the communication distance. Accordingly, this standard needs to be a core stream standard within the IEEE 802.11 series for IoT devices as the participating devices in IoT require a long communication distance rather than of high throughput performance. 
\end{itemize}

\subsubsection{\textbf{WiMAX}}

IEEE 802.16 is a collection of wireless broadband standards that provides data rates from 1.5 Mb/s to 1 Gb/s. The recent update (802.16 m) provides data rate of 100 Mb/s for mobile stations and 1 Gb/s for fixed stations. The specifications are readily available on the IEEE 802.16 working group website (IEEE 802.16, 2014).

\subsubsection{\textbf{Wireless Personal Area Network/Wireless Neighborhood Area Network}}

\begin{enumerate}
\item \textbf{Alliance}:
There exists alliance groups that develop products for IoT due to the reason that, barely, all IEEE standards for wireless communication define fundamental specifications and not to the extend to make products with mutual compatibility.
\textbf{Wireless HART} \cite{song2008wirelesshart}, \textbf{Thread} \cite{keythread}, \textbf{Wi-Sun} \cite{beecher2013wi, beecher2016wi} and \textbf{ZigBee} \cite{lawson2014zigbee, links2015power} are some of the specifications that are based on the IEEE 802.15.4. 

\textit{- Wireless HART} is a wireless standard that expands HART (Highway Addressable Remote Transducer) standard of digital industrial automation protocols for processes automation in factories. One of the advantages of HART is that it has a backward compatibility to traditional HART instruments. 

\textit{- Thread} is an IPv6 based mesh topology network protocol that provides Thread networking stack on top of the IEEE 802.15.4; this will allow each Thread end device to connect to the Internet through native IP protocols. 

\textit{- Wi-Sun} stresses on field area networks for applications (e.g., home energy management, distribution automation and advanced metering infrastructure); it helps in provisioning secure IPv6 communications over an IEEE 802.15.4g based wireless mesh topology network. 

\textit{- ZigBee}'s latest version, 3.0, provides a seamless interoperability among wide range of smart IoT devices. It also defines standard specifications to all levels of network specially applications levels for practical services.

\textbf{IEEE 802.1 Time Sensitive Networking} \cite{keytsn} aims at realizing deterministic communication that ensures real-time communication, transmission delay and data throughput. The main advantage of this standard is to realize both real-time and critical message deliveries on standardized Ethernet components. This standard is an extension protocol of Ethernet AVB protocol (IEEE 802.1 Audio/Video Bridging) in which those extensions provide minimal transmission latency and high availability in comparison to traditional wireless standards.

\textbf{CSRmesh}, protocol runs over Bluetooth Smart \cite{gomez2012overview}, provides message relying over a number of Bluetooth Smart devices as well as enables different products, such as tablets, smartphones, etc. to employ Bluetooth Smart to interact with devices within the range of the CSRmesh network directly.

\textbf{Z-Wave} \cite{alliance2015z} listed as wireless communication protocol on sub-GHz band that is used for home automation; it provides reliable, low-latency transmission of data packets chunks at speeds up to 100Kbps. The physical and MAC layers of this standard use source-routed mesh network architecture to help delivering messages to the indented destination. Such usage complies with ITU-T G.9959 recommendations \cite{keyITUT}. Like typical standards, Z-Wave network is identified via a Home ID where each participating device is identified via a Node ID as well. In general, Z-Wave has two basic device types: (a) controllers to control other Z-Wave devices, and (b) slaves which are controlled by other Z-Wave controllers. In addition, the Z-Wave alliance defines some profiles specifically for home automation, e.g., ZigBee 3.0, hence guaranteeing the interoperability factor between devices of different vendors.

\item \textbf{IEEE 802.15.4}:
This standard is allocated to low-rate wireless personal area networks (LR-WPANs) \cite{gutierrez2004low}, thus, it primarily focuses on both low transmission rate (250 Kbps) as well as power consumption as the participating IoT devices are assumed to operate in a tiny/limited battery capacity. There exist three types of nodes in such standard: (a) Reduced-function devices 'RFDs', (b) Full-function device 'FFD', and (c) PAN coordinator. Such nodes can coordinate to construct either peer-to-peer or star topology networks, but since the limitations that are associated with the smart devices, the maximum packet size is limited to 127 Bytes that include both a header as well as a payload to be transmitted. 

\begin{enumerate}
\item \textit{IEEE 802.15.4g} another standard that is an extension of the IEEE 802.15.4 for low-rate wireless neighborhood area network. It facilitates the control process of the very large scale applications, such as Smart Utility Networks (SUNs) with minimal infrastructure, in the presence of many fixed end-nodes. Metering of electricity, gas, etc., are the targeted utility out of this extended standard. In general, IEEE 802.15.4g builds a smart WNAN; it extends a data payload size of 2,048 Bytes compared to the traditional IEEE 802.15.4. The IEEE 802.15.4e is a standard for MAC layer mechanisms; it is used to realize a low power intermittent operations. It is used alongside with IEEE 802.15.4g as it does not define any physical layer in its configuration.  

\item \textit{IEEE 802.15.4k} is physical layer specification for low energy critical infrastructure monitoring network. It is used to provide a low transmission rate (< 10Kbps) as well as a long distance communication (> 1km). As a consequence, the IEEE 802.15.4 will be from the mainstream standards that are going to be used for the IoT devices. 
\end{enumerate}
\end{enumerate}

\subsubsection{\textbf{Wireless Wide Area Network}}

IoT services are incorporating Cloud services to provide functions to the end-IoT nodes. Hence, communication technologies for wide area networks (WANs) are crucial to realize practical services for IoT. Below is a discussion of the main stream for WANs.

\begin{itemize}
\item \textbf{Third Generation Partnership Project} \cite{unknown20103rd} already developed a number of standards dedicated for cellular network systems (i.e., 2G/3G/4G/5G mobile communication). The main goal of this standard is to achieve high-speed communication to let smart devices communication over cellular networks and exchange data in a quick manner (where data rates range from 9.6 Kb/s (2G) to 100 Mb/s and 1Gb/s (4G/5G)). Second generation (2G) includes GSM and CDMA, third generation (3G) includes UMTS and CDMA2000 and the fourth generation (4G) includes LTE. Trends for 4G and 5G cellular networks have two main categories: (a) high-speed communication, and (b) low-speed communication associated with low power consumption. For example, LTE Advanced Release 13 is shedding the light on the new specifications for the IoT devices as well as defining new terminal categories, such as category M1 which is designed to support a narrow band communication and Narrow Band (NB)-IoT which limits a bandwidth that is less than 180 KHz. Furthermore, it does update the Power Saving Mode 'PSM' specifications and does define an extended Discontinuous Reception 'DRX' to prolongate an intermittent reception interval of a paging mechanisms to drop the power consumed.

\item \textbf{Low-Power Wide-Area Networks} (LPWAN) is a new designed standard for wireless communication technology; it supports low data rate, low power consumption and long-distance communication. As a concrete example of this standard, we highlight LoRa \cite{augustin2016study} and SIGFOX \cite{SIGFOX} below:

- \textit{LoRa} is intended for wireless battery operated smart \textit{things} to enable IoT. It has a star topology where base stations are seen as a transparent bridge that relay messages between endpoints and servers. Data rates in LoRa are ranging from 0.3 Kb/s to 50 Kb/s and it operates in 868 and 900 MHz ISM bands. A base station of LoRa is not that expensive in comparison to that of \textit{SIGFOX} because radio devices for an endpoints and base stations almost have the same specifications. Mainly it aims at guaranteeing interoperability between different operators in one open global standard. In general, LoRa provides symmetric links for the endpoints available and helps attached \textit{things} to have battery life up to 10 years.

- \textit{SIGFOX} builds cellular style systems to serve communication services while employing an ultra-narrow band (UNB) technology, hence, it helps network operators to adopt their technology for conventional IoT deployments. SIGFOX helps vendors to develop their own products via the endpoint available; endpoints in this technology use bidirectional communication to provide high quality communication service. \newline 

\item \textbf{GSMA/eSIM} \space
The specifications for GSMA Embedded SIM (eSIM) \cite{specification2014provisioning} provide standard mechanisms for M2M connection management. Conventional cellular devices need a physical traditional SIM card to connect to network operator; SIM card should be installed into device's slot to function and connect to the networks available. GSMA/eSIM assume Embedded Universal Integrated Circuit Card (eUICC) as a new embedded SIM function for smart devices; the eUICC identification information (eICCiD) associated to the eUICC allows both the over the air `OTA' provisioning of an initial operator subscription as well as changing of subscription from one operator to another. An eSIM selects a profile from the installed profiles according to the commands from the subscription manager of a mobile network operator. 
\end{itemize}

\subsubsection{\textbf{\textit{OASIS}}} 

The technical committee of OASIS has published the Extensible Resource Identifier (XRI) standards and the Extensible Resource Descriptor (XRD) \cite{wachobextensible, hammer2009extensible}. The syntax of the XRI leverages both the uniform resource identifier (URI) and the internationalized resource identifier (IRI) specifications \cite{durst2004internationalized}; it defines a generic syntax for structuring abstract identifiers so that they can be shared among various application domains and embedded transparently found within different URI schemes. Thus, the XRI provides a standard mechanism for resources identification in an abstract way and independently of its concrete representation. Whereas, the XRI Resolution Standard defines a generalized secure protocol for resolving XRI information by the means of the resource descriptions and the HTTP/HTTPS URI information \cite{wachobextensible}. 

\subsection{Internet of Things Platforms}

In IoT systems, messages can travel from one end-device to another application and/or device via the available WAN, PAN and platform layer. Recently, a number of platform layer standards have been suggested in some organizations to address fast delivery of messages between participating smart nodes and devices. 

\begin{itemize}
\item \textbf{OneM2M} standard provides a common M2M service layer that can be easily embedded within different hardware and software components \cite{key2012}; it defines a number of use cases and requirements for a common set of protocols, APIs, identification and naming of smart devices and applications, security and privacy mechanisms, interoperability, information model and data management, management aspects, as well as services. The benefit of the OneM2M standard is that it considers horizontal service domains in IoT to help reusing information and creating new values out of such reused information as well.

\item \textbf{\textit{Web of Things (WoT)}} is a new standard that is used to handle real-world objects and help them be a part of the World Wide Web \cite{key2007_1}; it provides a simplified application layer to create new IoT products. Despite the fact that this standard uses HTML5/JavaScript as developing languages, the developed codes should be able to operate on different kind of hardware, software and operating system components to realize such integration. Moreover, WoT defines additional standards to obtain information via Web APIs, as a consequence, web applications on every IoT smart device are able to connect each other via those APIs.

\item \textbf{IEEE 2413} defines an architectural framework for IoT where it includes descriptions of different IoT domains, definition of IoT domain abstraction and identification of commonalities between different considered IoT domains \cite{key2014_1}. Furthermore, it provides some reference architectures to build a reference model according to a practical service application. The architectural framework defined in this standards focuses on cross-domain interaction, aids system interoperability, functional compatibility as well as fuels the growth of the IoT market.
\end{itemize}

\subsection{System Model}

Recent IoT services have been developed on a vertical system model in which each layer has been designed by an organization or a company. In turn, a recent trend of standardization considers horizontal system model to achieve scalable and interoprable operations in IoT's services \cite{naito2017survey}. As the previous sections indicate, different types of standards protocols have been suggested to establish a cooperative mechanisms between horizontal services domain; those protocols usually focus on a standardization in an applications layer. Hence, they assume that inter-accessibility between nodes is guarantee. Contrarily, practical IP networks have some issues with inter-accessibility because of the differences of IP protocol versions and firewalls. Thus, the proposal of a new IoT service layer design to realize inter-operation between end-nodes in different networks is needed.

The current IoT layer model assumes that IP networks are transparent which means end-nodes have the ability to access each other \cite{naito2017survey}. Such assumption would be reasonable when an IoT service is operated in a close IP network, e.g., smart metering systems. As a consequence, a new IoT layer model should be considered to have a middleware layer between the IP and transport layers to achieve a transparent connectivity between those end-nodes.

\section{Internet of Things Integration Practical Guidelines} 
\label{state of practice}

A framework for sustainable interoperability in IoT is needed; this framework can (and should) learn from the best-of-breed interoperability solutions from related domains to take the good approaches while understanding the differences that the IoT poses \cite{key2017_2}. The below five steps can address some of the most common challenge in IoT integration: 

\begin{enumerate}
\item \textit{Adopt an API-first approach}: this is particularly relevant to IoT projects as they rely on mobile and Cloud Computing technologies that already use an API-centric approaches. This, however, should not be misinterpreted  as an API-only approach; API alone is not sufficient to address all the capabilities needed to securely and reliably scale up integration in large scaled distributed systems \cite{key2017APIs}. 

\item \textit{Communication requirements identification for IoT devices}: first identify how \textit{things} are going to communicate then select the best technology accordingly (i.e., ranging from cellular networks to short range wireless technologies as Bluetooth or ZigBee) \cite{tao2014cciot,wang2014iot}. It is also vital to consider other factors, such as the number and type of \textit{things}, and how different technologies will handle such variables. Then identify the best network topology "considering new trends as Fog edge computing or gateways" that best suits the requirements for devices' autonomy, aggregation, localized computing, etc. Once such areas are qualified, it will be easier to assess whether a bundled IoT platform meets the requirements for the integration or whether additional solutions would be needed to build a network for \textit{things}.

\item \textit{Leveraging Cloud for data and process integration}: This step focuses on IoT platforms integration with core business processes \cite{villari2016leveraging}. The built-in integration capabilities of IoT platforms are good enough for initial deployments. For example, using an IoT platform for initial implementation, then use a commercial integration solution, e.g., iPaaS \footnote{see source: www.gartner.com/it-glossary/information-platform-as-a-service-ipaas/} platform, to scale up the project being worked on, support more complex integration to implement work flows or to access advanced integration features, e.g., high performance, general purpose translation. 

\item \textit{Using selective traditional software}: Most enterprises have substantial investments towards on-premises integration middlewares. Despite the non-optimality of those tools for IoT devices connectivity or Cloud services integration, they can help if the used IoT platform must integrate with data and applications that are mostly on-premises \cite{key2017interoperabilityChallenge}. 

\item \textit{API management tools usage}: The capabilities of API change in different IoT platforms or other types of middleware. A good management of API involves adding a third-party API management solution to IoT platforms to ensure secure and reliable scaling as APIs widely increase \cite{key2017APIManagementTools}. This holds true if projects involve many APIs are exposed to public networks to provide sensitive or restricted data to end-users. 
\end{enumerate}

\section{Internet of Things Software Development and Challenges}
\label{software}

Emergence of IoT brings a new class of software and--or applications and additional efficiency constraints due to the limited resources of the \textit{things}.
Software-specific requirements, communication and connectivity ability of devices have introduced new challenges for IoT systems' interoperability. With the ever increasing number of interconnected embedded devices, there is a need for new software solutions to help developers manage software and hardware interoperability issues in a scalable, smart and an efficient way.

\subsection{Internet of Things Proposed Software Architectures}

IoT architecture is still under construction, it is not fixed and does not have a concrete shape yet. However, rapid development of IoT has triggered a wave of unreasonable expectations \cite{ning2011future}. For example, some industries have launched huge projects despite that the key technologies, including the basic architecture of IoT, are still not fully determined. Hence, it would be dis-advantageous to IoT development and may cause an unexpected loss.

A number of researcher adopted microservices architecture \cite{dmitry2014micro} which is a new software design pattern that aims at addressing interoperability issues by promoting lightweight--independent services that perform single functions and collaborate with other similar services using well-defined interfaces. 
Each microservices is dedicated for a single functionality, hence, independent services can be easily deployed into the production environment and any service modification would not affect the whole system. 
A microservices architecture has multiple advantages, such as independent deployment, complexity under control, providing more options for technology stack and fault tolerance. All these desirable properties facilitate the development of IoT software and applications on large scale. Furthermore, they help standardize services' interfaces allowing them to communicate with each other even if they have been developed and deployed on heterogeneous platforms.

\subsection{Real-world IoT Deployment Platforms}

Eclipse Hono, which originates from Eclipse IoT project \cite{whitepaper2}, allows the provisioning of remote service interfaces for connecting different devices and interacts with them uniformly regardless of their type or communication protocol. The platform provides a number of protocol implementations, i.e., HTTP REST, MQTT, etc., Cloud front/back-ends, Machine-to-Machine (M2M) management, devices’ authentication and management among others. It also provides the possibility of multi-tenancy, meaning that the same infrastructure can be shared between different tenants to allow the scaling options for the development of huge software platforms and applications.

In addition, Kuksa \cite{whitepaper3}, an open-source platform, addresses specific demands of the connected devices where it uses and extends existing technologies to ease development, analysis and activities for IoT and Cloud-based approaches for interconnected objects. It also provides a basis for new application fields as it contains a Cloud platform that interconnects a wide range of devices via Internet connections. This platform is supported by an integrated open source software development environment including various technologies to cope with software challenges for devices collaborating in the IoT system.

\section{Future Directions for the Internet of Things} 
\label{future}

After connecting people anytime and everywhere, the next important step is to integrate heterogeneous \textit{things} among themselves and with the Internet \cite{vermesan2013internet}. This integration will allow the creation of value-added interoperable services and applications, enabled by their interconnections, in a way that they can be integrated with the current and new business and development processes. 
 
\paragraph{\textbf{Edge Computing}} introducing IoT data sources globally strengthens challenges already faced with ``Big Data'' \cite{data2013better}, in particular when considering the typical deployment models of the IoT where data provided by smart sensors (i.e., at the edge of the infrastructure) are transmitted to data centers (i.e., at the core of the infrastructure) for processing. Conveying entire data sets across infrastructures becomes an unrealistic proposition, where instead, approaches that strive to collect data and do computations closer to the sensors (e.g., Edge Computing \cite{bonomi2012fog}, Near-Data processing \cite{balasubramonian2014near}, etc.) are more practical alternatives to such scenarios. Edge Computing disseminates data to be processed away from the core of the infrastructure and closer to the latter's edge, as close to the data sources as possible, even trying to make such processing on/at the device itself. Placing the processing near to data sources is beneficial in cases of video, for example, whose transport across infrastructure can claim considerable network resources (i.e., not forgetting the resources that are needed for its storage). It is considered to be more resource-effective to process real-time information at its source than extract all relevant features, such as objects, faces, etc.\ in the Cloud. With Big Data, the cost of transmitting data from its source to the computing facilities is a major concern. Reducing data movement by making computations near to data sources is an approach that is known as Near-Data Processing (NDP) \cite{balasubramonian2014near}. Edge computing converts the communication protocols used by the participating devices into a language that modern smart \textit{things} can understand; this makes it easier to connect \textit{things} with modern IoT platforms.

\paragraph{\textbf{Integrating Social Networking with IoT}} Social Internet of Things (SIoT) has been suggested by Atzori et al.\ \cite{atzori2012social} to address the strong interest of using social networking to enhance the communications among different IoT things. There is a trend of moving from IoT to a new vision of the WoT that allows smart objects to become active actors and peers on the Web \cite{atzori2014smart,guinard2011internet}. Social networks are proposed to perform automatically the discovery of things and services and, thus, improve the scalability of IoT similar to human social networks. 

\paragraph{\textbf{Developing Context-Aware Middleware Solutions for IoT}} By 2020, billion of things will be connected together. When such a huge number of things is connected to the Internet, it will not feasible for individuals to process all the data collected by them. Context-awareness computing techniques, e.g., middleware solutions for IoT, are suggested to understand sensor data in better ways to help decide which data must be aggregated and processed \cite{wang2013application}. Currently, most middleware solutions do not possess context-awareness capabilities. The European Union has identified that context awareness is a crucial research area and specified a time-frame (2015-2020) for context-aware IoT computing R\&D \cite{vermesan2011internet}. Middleware solutions deals with heterogeneous devices and manages interoperability among them by understanding the sensory data collected besides providing support to process and store those data and make their interpretation easy.

\paragraph{\textbf{Internet of Nano-Things}} Another vision that involves integrating even more devices into the IoT is the Internet of Nano-Things. The Internet of Nano-Things is viewed as the interconnection of nano-scale devices via the Internet and communication networks. While those devices are purposed to communicate via electromagnetic communications, there are a huge number of technical challenges that should be addressed before such idea becomes feasible \cite{akyildiz2010internet}. The Internet of Nano-Things is considered a more granular approach to ubiquitous computing than the conventional IoT by embedding nano-sensors inside the devices to communicate together through nano-networks via the Internet for global connection among devices around the world.  

\section{Conclusion}
\label{conclusion}

This paper presented a review on the integration and interoperability issues in IoT and some proposed solutions. Since its inception, IoT services and applications have been developed to be adapted in a vertical service model, i.e., in which every layer has been designed by a company or organization. Interoperability and fragmentation among various service domains are a major challenge. As a consequence, the latest trends of IoT technologies are to adapt horizontal service domain to achieve interoperability between participating \textit{things}. Various standards and protocols have been proposed by a number of IoT consortiums to tackle the integration issues. However, current IoT technologies do not fully make an inter-operation between various devices in different networks. With the help of IP mobility technologies, interoperability between devices in different networks has been realized. As the IoT market develops, interoperability will be of a crucial factor to the commercial success of IoT services and applications to enable Internet-based collaborative technologies, hence, knowledge and understanding of the IoT standardization landscape and the established architectures for the IoT paradigm is essential.

In future work, we will further develop our SLR in two directions. First, we will include more papers to cover more IoT devices, protocols, and standards, in particular networking protocols. Second, we will define and assess various aspects of the IoT devices, in particular security, energy consumption, and usability in addition to interoperability. Future work also includes developing means for users and developers of IoT devices to assess possible integration issues before using/releasing the devices into action.

\bibliographystyle{IEEEtran}
\bibliography{Bibliography}

\begin{IEEEbiographynophoto}{Mohab Aly}
is a former Network Specialist at TELUS Telecommunications Incorporate in Canada and a Technical Solutions Representative for Google for its Google Cloud Platform (GCP). He received his B.Sc. in Computer Science and IT from Ahram Canadian University (ACU) in 2010, and M.A.Sc. in Computer Engineering from \' Ecole Polytechnique de Montr\' eal in Canada in 2013. He is currently pursuing a Ph.D. degree in Industrial Engineering with a specialization of Computer Integration and Manufacturing at \' Ecole Polytechnique de Montr\' eal as well. His research interests include Internet of Things, Cloud Computing, Computer Networking, Parallel Computing and Industrial Engineering. He has published several papers in international conferences, including IWCMC, ICIN and RAMS.
\end{IEEEbiographynophoto}

\begin{IEEEbiographynophoto}{Foutse Khomh}
is an associate professor at Polytechnique Montr\' eal, where he heads the SWAT Lab on software analytics and cloud engineering research (http://swat.polymtl.ca/). He received a Ph.D in Software Engineering from the University of Montreal in 2010, with the Award of Excellence. His research interests include software maintenance and evolution, cloud engineering, service-centric software engineering, empirical software engineering, and software analytic. He has published several papers in international conferences and journals, including ICSM(E), ASE, ISSRE, SANER, ICWS, HPCC, IPCCC, JSS, ASEJ, JSEP, EMSE, and TSE. His work has received a ten-year Most Influential Paper (MIP) Award, three Best Paper Awards, and fifteen nominations for Best paper Awards. He has served on the program committees of several international conferences including ICSM(E), SANER, MSR, ICPC, SCAM, ESEM and has reviewed for top international journals such as SQJ, JSS, EMSE, TSE and TOSEM. He is on the Review Board of EMSE. He is program chair for Satellite Events at SANER 2015, program co-chair of SCAM 2015, ICSME 2018, and ICPC 2019, and general chair of ICPC 2018. He is one of the organizers of the RELENG workshop series (http://releng.polymtl.ca) and has been guest editor for special issues in the IEEE Software magazine and JSEP.
\end{IEEEbiographynophoto}

\begin{IEEEbiographynophoto}{Yann-Ga\"{e}l Gu\'{e}h\'{e}neuc}
is full professor at the Department of Computer Science and Software Engineering of Concordia University since 2017, where he leads the Ptidej team on evaluating and enhancing the quality of the software systems, focusing on the Internet of Things and researching new theories, methods, and tools to understand, evaluate, and improve the development, release, testing, and security of such systems. Prior, he was faculty member at Polytechnique Montr\' eal and Universit\' e de Montr\' eal, where he started as assistant professor in 2003. In 2014, he was awarded the NSERC Research Chair Tier II on Patterns in Mixed-language Systems. In 2013-2014, he visited KAIST, Yonsei U., and Seoul National University, in Korea, as well as the National Institute of Informatics, in Japan, during his sabbatical year. In 2010, he became IEEE Senior Member. In 2009, he obtained the NSERC Research Chair Tier II on Software Patterns and Patterns of Software. In 2003, he received a Ph.D. in Software Engineering from University of Nantes, France, under Professor Pierre Cointe’s supervision. His Ph.D. thesis was funded by Object Technology International, Inc. (now IBM Ottawa Labs.), where he worked in 1999 and 2000. In 1998, he graduated as engineer from \' Ecole des Mines of Nantes. His research interests are program understanding and program quality, in particular through the use and the identification of recurring patterns. He was the first to use explanation-based constraint programming in the context of software engineering to identify occurrences of patterns. He is interested also in empirical software engineering; he uses eye-trackers to understand and to develop theories about program comprehension. He has published papers in international conferences and journals, including IEEE TSE, Springer EMSE, ACM/IEEE ICSE, IEEE ICSME, and IEEE SANER. He was the program co-chair and general chair of several events, including IEEE SANER’15, APSEC’14, and IEEE ICSM’13.
\end{IEEEbiographynophoto}

\begin{IEEEbiographynophoto}{Hironori Washizaki}
is the director and professor at the Global Software Engineering Laboratory (http://www.washi.cs.waseda.ac.jp/), Waseda University, Japan. He also works at National Institute of Informatics as the visiting professor, and at SYSTEM INFORMATION CO., LTD. and eXmotion Co., Ltd. as outside directors. He visited \' Ecole Polytechnique de Montr\' eal in 2015 during his sabbatical year. He obtained his Doctor’s degree in Information and Computer Science from Waseda University in 2003. He has been leading various projects including Goal-Oriented Quantitative Management, Cloud Security and Privacy Metamodel, TraceANY for tracing anything, Waseda Software Quality Benchmark, enPiT-Pro SmartSE (https://smartse.jp/) as the professional education program, and G7 Programming Learning Summit (http://g7programming.jp/) as the learning environments study. He has contributed to many societies such as IEEE Computer Society Ad Hoc Committee for Program Boards Integration Chair, ISO/IEC/JTC1 SC7/WG20 Convenor, IPSJ SamurAI Coding Director, International Journal of Agile and Extreme Software Development Editor-in-Chief, IEEE ICST 2017 PC Co-Chair, IEEE CSEE\&T 2017 PC Co-Chair, APSEC 2018 PC Co-Chair, IEEE COMPSAC 2018 Local Chair and AsianPLoP 2018 General Chair. 
\end{IEEEbiographynophoto}

\newpage
\begin{IEEEbiographynophoto}{Soumaya Yacout}
is a Full Professor in the Department of Mathematics and Industrial Engineering at \' Ecole Polytechnique de Montr\' eal, Canada (http://www.polymtl.ca/expertises/en/yacout-soumaya). She holds a D.Sc. in Operations Research from Georges Washington University in USA, a B.Sc. and a M.Sc. in Industrial Engineering from Cairo University in Egypt. Her research interests include Condition Based Maintenance, distributed decision making for product quality and Industry 4.0. She is a senior member of the American Society for Quality and a member of the Institute of Industrial Engineering (IIE) and the Canadian Operational Research Society.
\end{IEEEbiographynophoto}

\end{document}